\documentclass[onecolumn,superscriptaddress]{article}

\usepackage{graphicx}
\usepackage{epsfig}
\usepackage{dcolumn}
\usepackage{bm}
\usepackage{amsmath}
\usepackage{amssymb}
\usepackage[dvips]{hyperref}
\usepackage[scale=0.82]{geometry}
\hypersetup{colorlinks=true,linkcolor=blue,citecolor=blue,urlcolor=black}

\newcommand{\ket}[1]{\ensuremath{|\,{#1}\,\rangle}}
\newcommand{\bra}[1]{\ensuremath{\langle\,{#1}\,|}}

\begin{document}

\title{\vspace{-1.5cm}\bf Quantum key distribution session with 16-dimensional photonic states}
\author{S. Etcheverry,$^{1,2}$ G.~Ca\~nas,$^{1,2,3}$ E.~S.~G\'omez,$^{1,2,3}$ W.~A.~T.~Nogueira,$^{1,2,3}$ \\ C.~Saavedra,$^{1,2}$ G.~B.~Xavier,$^{2,3,4}$ and G.~Lima$^{1,2,3,*}$ \vspace{0.1cm}\\ \normalsize\it
$^1$Departamento de F\'isica, Universidad de Concepci\'on, 160-C Concepci\'on, Chile\\ \normalsize\it
$^2$Center for Optics and Photonics, Universidad de Concepci\'on, Concepci\'on, Chile\\ \normalsize\it
$^3$MSI-Nucleus for Advanced Optics, Universidad de Concepci\'on, Concepci\'on, Chile\\ \normalsize\it
$^4$Departamento de Ingenier\'ia El\'ectrica, Universidad de Concepci\'on, 160-C Concepci\'on, Chile \\ \normalsize\it
$^*$e-mail: glima@udec.cl}

\date{\normalsize Dated: \today}

\maketitle

\textbf{The secure transfer of information is an important problem in modern telecommunications.
Quantum key distribution (QKD) provides a solution to this problem by
using individual quantum systems to generate correlated bits between remote parties, that can be used to extract a secret key.
QKD with $D$-dimensional quantum channels provides security advantages that grow with increasing $D$. However,
the vast majority of QKD implementations has been restricted to
two dimensions. Here we demonstrate the feasibility of using higher
dimensions for real-world quantum cryptography by performing, for the
first time, a fully automated QKD session based on the BB84 protocol with $16$-dimensional quantum states. Information is
encoded in the single-photon transverse momentum and the required states are dynamically generated with programmable spatial light
modulators. Our setup paves the way for future developments in the field of experimental high-dimensional QKD.}

\bigskip

Theft of electronic information is a major problem in modern telecommunications. For instance, recent advances in computing technology such as the ever growing parallel processing power of graphical processing units (GPUs), have allowed the brute-force cracking of passwords to be greatly sped up \cite{GPU1, GPU2}.
A very promising solution to the issue of information security is quantum key distribution (QKD), also called quantum cryptography \cite{Gisin_RMP}. Quantum cryptographic schemes allow remote parties to share a secret key (a sequence of bits) through the use of quantum channels. This key is later used for secure information transmission, and the advantage of QKD is that the presence of an eavesdropper may, in principle, be detected. The first QKD protocol was proposed in 1984 by Bennett and Brassard, called BB84 \cite{BB84}. It employs two sets of orthogonal quantum states spanning two mutually unbiased bases (MUBs) \cite{Wooters2}. BB84 is a prepare-and-measure QKD protocol, in which single quantum systems are exchanged between the communicating parties (Alice and Bob) for the establishment of each shared secret key bit.

Most experimental implementations of quantum cryptography have been done with BB84-based QKD schemes. Experiments with increased performance (i.e. higher transmission distance, larger secure key generation rate, more compact schemes, etc...) have been demonstrated in the past few years \cite{Gobby_APL_2004, Takesue_natphoton_2007, Stucki_NJP_2009, Guix_NJP_2009, Shields_APL_2010, Guo_OL_2012}, and the vast majority of these demonstrations have employed only two-dimensional quantum systems. The use of high dimensional systems has been nevertheless widely deployed in classical digital communications for many years, as a way to maximize the efficiency of the transmission channel \cite{Proakis_book}. With the same goal, the first studies on the theoretical framework to generalize the BB84 protocol to high dimensional state spaces were carried out over a decade ago \cite{Tittel_PRA_2000, Boure_PRA_2001, Cerf_PRL_2002}. It was shown that there is an important additional feature in moving to higher dimensions in QKD: increased security, which is alone a major motivation to implement high-dimensional quantum cryptography. Nonetheless, due to experimental challenges, there have been only proof-of-principle experiments of QKD in higher dimensions \cite{Zeilinger_NJP_2006, Howell_PRL_2007, Steve_PRL_2006}.

Recently, on the theoretical front, the interest on high-dimensional QKD did not wane as security bounds were derived considering finite key lengths \cite{Sheridan_PRA_2010}. Moreover, an important experimental tool has emerged in recent years to dynamically manipulate light in free-space, the spatial light modulator (SLM) \cite{Twi1}. The capabilities of the SLMs have been explored in experiments of quantum information and it is now widely recognized as a tool for the manipulation of high-dimensional quantum systems (qudits) encoded in the transverse momentum of single photons \cite{Padgett_2006,Glima_opex_2009, Padgett_nature_phys,Torres_nature_phys,Glima_opex_2011,SPadua_opex}. In classical optical communications, SLMs have also been used to multiplex and demultiplex classical high-speed data streams, further increasing the channel capacity \cite{Willner_nat_photon}. Nevertheless, in spite of all recent progress, a demonstration of a quantum key exchange session using the BB84 protocol in higher dimensions is still lacking. In this paper this issue is finally addressed.

We report an automated QKD session between Alice and Bob, based on the BB84 protocol extended to $16$-dimensional quantum systems \cite{Cerf_PRL_2002}. We employ the linear transverse momentum of single-photons as the degree of freedom for encoding the qudit states \cite{Neves_PRL_2005}. The single-photons are produced by Alice from attenuated laser pulses. At Alice and Bob's sites, the quantum states spanning the MUBs are randomly produced with the help of SLMs, dynamically introducing relative phases between the paths available for the propagation of the single photons \cite{Glima_opex_2011}. The entire setup is synchronized working at a repetition rate of 30 Hz, which is the current physical limit of standard commercially available SLMs. The custom electronics required for the full automated execution of the protocol were developed on field programmable gate array (FPGA) units. The measured quantum bit error rate (QBER) of the correlated bits generated was smaller than half of the bound allowed for extracting a secret key while considering the general coherent attacks; and close to three times smaller than the limit for individual attacks \cite{Cerf_PRL_2002,Sheridan_PRA_2010}.

\section*{\label{sec:level1}Results}

In this demonstration we take advantage of the flexibility allowed by SLMs to encode information using the transverse wave profile of single photons. Although it is a continuous degree of freedom, it can be discretized using slits to define different possible paths for the photon transmission \cite{Neves_PRL_2005}. The SLMs are built with liquid crystal displays (LCDs) and require the use of polarizers. They allow an individual control of the gray level of each pixel on the LCD \cite{Moreno_opex}. While changing the pixel gray level, the polarization is changed and when combined with the polarizers, the light transmissivity of the pixel is varied. Together with this effect a relative phase change may also be induced in each pixel. The slits can be directly generated on the SLM. This is done by programming the gray levels of all the LCD pixels such that areas with high transmissivity correspond to the slits, and everywhere else has zero transmission. By employing a set of $D$ parallel slits a Hilbert space of $D$ dimensions is constructed. The state after the slits can be written as the following coherent superposition:
\begin{equation} \label{State}
\ket{\Psi} = \frac{1}{\sqrt{D}}\sum_{l=-l_D}^{l_D} \ket{l}e^{i\phi_l},
\end{equation} where $l_D=(D-1)/2$ . $\ket{l}$ represents the state of the photon transmitted by the lth-slit of the SLM \cite{Neves_PRL_2005}. The uniform amplitude distribution is obtained with an adequate choice of the spatial beam profile on the plane of the slits \cite{Glima_opex_2011}. This state can be generated provided that the single-photons have a larger transverse coherence length than the distance between the outermost slits. $\phi_l$ are the relative phase shifts at each slit. Each $\phi_l$ can be dynamically changed in real-time in the SLMs.

In standard BB84 quantum cryptography, the required states are two vectors of two-dimensional MUBs, with a total of four states required. In the case of a multi-dimensional BB84 protocol, the individual states to be sent and measured $\{ \mid \Psi^{(\beta)}_0 \rangle, \dots , \mid \Psi^{(\beta)}_{D-1} \rangle \}$ belong to two $D$-dimensional MUBs. The index $\beta$ denotes a specific MUB with $\beta= \alpha, \alpha'$. In this experiment $16$ slits are employed, generating a $16$-dimensional Hilbert space. Consequently $32$ states had to be generated, $16$ for each MUB. Of the large number of families of MUBs that exist in 16 dimensions \cite{Klimov}, we chose two MUBs that only required phase modulations of $0$ and $\pi$ and no amplitude modulation (see Methods). This eased the implementation of the MUBs states with the SLMs.

\begin{figure}[t]
\centerline{\includegraphics[width=87mm]{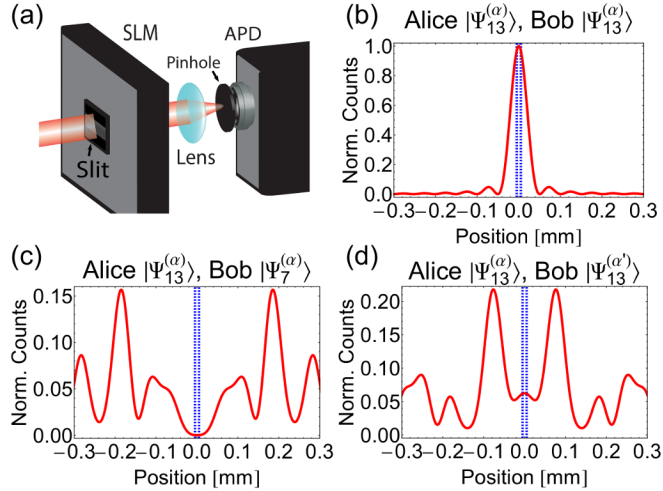}}
\caption{Detection procedure. (a) Detection apparatus. See the main text for details. (b) Numerical simulation showing the single-photon spatial distribution in the case that Alice and Bob have both chosen the same state $\ket{\Psi^{(\alpha)}_{k}}$ in the same MUB $\alpha$. We have arbitrarily chosen states 13 from MUBs $\alpha$ and $\alpha^{'}$ and state 7 from $\alpha$ to illustrate the distributions. The dashed blue dotted lines indicate the boundaries of the 10$\mu$m circular pinhole used in front of the detector in the experiment. In this case, the probability of detecting the single-photon is maximum. (c) Alice and Bob choose different states $\ket{\Psi^{(\alpha)}_{k}}$ and $\ket{\Psi^{(\alpha)}_{k'}}$ from the same MUB $\alpha$. In this case the detection probability is null. (d) The detection probability when two vectors are chosen, from the two distinct MUBs. In this case a detection probability of 1/16 is obtained.} \label{Fig:Fig1}
\end{figure}

Alice needs to randomly choose in which MUB, $\alpha$ or $\alpha'$, she will encode her qudit. She then randomly chooses one of the $D$ possible states from this MUB. As we will explain next, in our scheme, this is done by using a phase-modulation SLM and pseudo-random subroutines to depict a sequence of pre-defined masks. Upon successfully receiving the photon, Bob needs to make a projective measurement. He also randomly chooses states from $\alpha$ and $\alpha'$. For this purpose, Bob uses a detection apparatus composed of a phase-modulation SLM, a lens and a ``point-like'' avalanche single photon detector (APD) [shown in Fig.~1(a)]. The single-photon detector remains fixed at the center of the focal plane. It covers the transverse blue zone of 10$\mu$m indicated in Fig.~1. At the SLM, Bob depicts phase-masks where the phases at each slit are chosen in accordance with the MUBs states given in the Methods section. In this case, the probability of single-photon detection is proportional to $| \bra{\Psi^{(\alpha)}_{k}}\Psi^{(\alpha)}_{k'} \rangle |^2$ ($|\bra{\Psi^{(\alpha)}_{k}}\Psi^{(\alpha')}_{k'} \rangle |^2$) when Alice and Bob choose the states belonging to the same (different) MUBs, where $k,k'=0,...,D-1$. Each time Bob matches his state choice with Alice's, while both choose the same MUB, constructive spatial interference occurs and the probability of single-photon detection (the single count rate) is maximal [See the example of Fig.~1(b)]. When the vectors are not matched, and Alice and Bob have chosen the same MUB, total destructive interference occurs at the transverse area of the detector [See the example of Fig.~1(c)]. When Alice and Bob's choices of MUBs are not the same (incompatible) the probability of detection is $1/D$ regardless of which states were chosen by both [See the example of Fig.~1(d)].

In our experiment the light source is a single-mode continuous-wave laser with a 690 nm emission wavelength (See Fig.~2). It is followed by an acousto-optic modulator (AOM), which sets the optical pulse repetition frequency. The width of the pulses depends on the average photon number employed. Two mean photon numbers per pulse were used, $\mu_a = 0.60$ and $\mu_b = 0.18$, to test the performance of our QKD prototype. A 50 ns optical pulse for the case of $\mu_a$ is generated by the AOM. For the results with $\mu_b$, a 20 ns optical pulse was used instead. An extra calibrated attenuator (not shown in Fig.~2) is also used to set precisely the desired average photon number at the output of Alice's station.  The average photon number was measured directly at Alice's output with a removable mirror and a single-photon detector with known overall detection efficiency (also not shown in Fig.~2).

We employed three SLMs, two located within Alice, and the other belonging to Bob. The first SLM (SLM1) is programmed to perform amplitude modulation in order to generate the slits. This stage defines the number of paths available for the photon transmission and, therefore, the dimension to be used in the experiment, 16 in our case. The mask projected on this SLM remains fixed throughout the experiment. The slits are 2 pixels wide, with a separation of 1 pixel between them, with each pixel having an area of 32 x 32 $\mu$m$^2$. The relation between the width of the slits and the separation between them was chosen to maximize the distribution probability in the center of the detection plane, when Alice and Bob take the same state [See Fig.~1(b)]. A set of polarizers is used before and after each SLM.

\begin{figure}[t]
\centerline{\includegraphics[width=87mm]{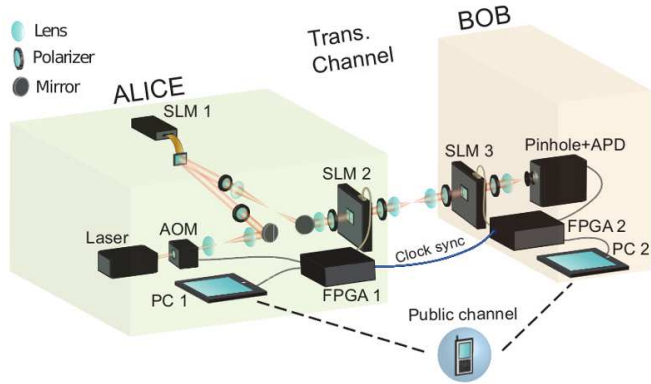}}
\caption{Experimental setup. The laser light source is followed by an acousto-optic modulator (AOM) which defines the optical pulses repetition frequency and width, see text for details. A FPGA unit is used by Alice to trigger the AOM and randomly project phase-masks in the SLM2 to encode the states required in the 16-dimensional BB84 protocol. Bob has a different FPGA unit to randomly perform the MUBs states projections and to count the single-photon detections at the avalanche photo-detector (APD). The FPGAs are connected between themselves for synchronization. Each FPGA is connected to a personal computer (PC1 and PC2) who perform the basis reconciliation procedure while communicating through an Ethernet cable, which acts as the public channel. The single-photons are transmitted from Alice to Bob through free-space with a telescope system forming the transmission channel (Trans. Channel).} \label{Fig:Fig2}
\end{figure}

A first set of lenses is used as a beam expander to illuminate the entire area of the slits at the SLM1 with attenuated pulses.  A second set of lenses forms the image of SLM1 onto the second SLM (SLM2). This SLM is configured for phase-modulation in order to implement the phase shifts required for constructing the MUBs states at Alice's station. This is done following the technique introduced in \cite{Glima_opex_2011}. The single-photon is then transmitted through a free-space channel (here consisting of a third telescope set of lenses), forming the image of Alice's output onto Bob's modulator (SLM3). This link is 50 cm long. A final lens following the SLM3 focuses the light onto the APD.

The entire system is synchronized at 30 Hz, with the master clock generated within Alice's FPGA and distributed to Bob through an electrical cable for synchronization. Alice's FPGA unit is used to trigger the AOM and generate the attenuated pulses. The pulses generated are naturally phase randomized due to the low repetition frequency of the pulses when compared to the laser coherence length. Nevertheless for higher operating frequencies, it can be actively phase randomized with schemes such as \cite{Lo_phase_random}. The bases and state choices done by Alice and Bob are performed independently using pseudo-random number generator subroutines in their individual FPGAs. For each clock pulse Alice and Bob choose, independently, a mask from the 32 distinct possibilities and project them onto SLM 2 and SLM3, respectively. Bob's FPGA unit records the single-photon detections at the APD. It also controls the detection window duration, which is 50ns when working with $\mu_a$ and 20ns for $\mu_b$. Each FPGA records the corresponding state choice and is connected to a personal computer (PC1 and PC2) who perform the basis reconciliation procedure through an Ethernet cable.

\begin{figure}[t]
\centerline{\includegraphics[width=87mm]{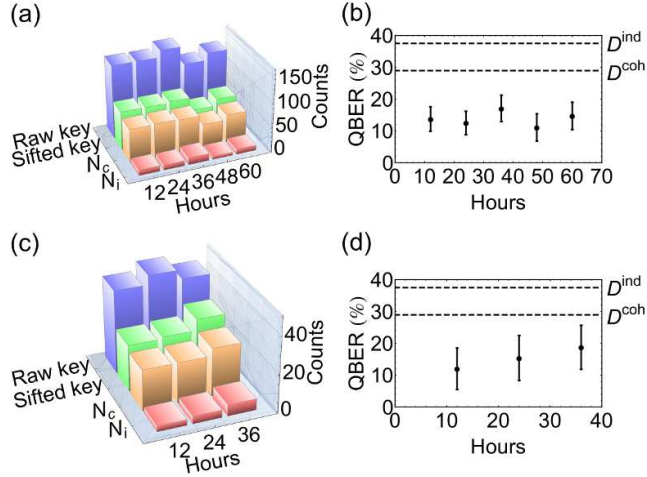}}
\caption{Experimental results. (a) Number of detections as a function of the number of elapsed hours for the QKD session with $\mu_a = 0.60$ photons per pulse. The detections are broken down into raw key (total detections), sifted key (sum of the detections when Alice and Bob choose the same MUB), $N_c$ (sum of all the detections when the same state from the same MUB is chosen by both Alice and Bob) and $N_i$ (sum of all the detections when different states from the same MUB are chosen). Each row sums up the results of 12 hours of continuous measurements. (b) The measured QBER as a function of continuous elapsed hours of the QKD session with $\mu_a$.  Error bars show the statistical error in the calculation of the QBER. The two security thresholds for general collective ($D^{\mathrm{coh}}$) and individual attacks ($D^{\mathrm{ind}}$) are also plotted with dotted lines. The average QBER is 13.4$\pm$4\% with an average sifted key rate of 27.0 bits / hour. (c) The same as part (a) but with $\mu_b$ = 0.18 photons per pulse. (d) Same as (b) but with $\mu_b$. The average QBER in this case is 15.6$\pm$7\% and an average sifted key rate of 11.1 bits / hour. Note, that in both cases the QBER is well below the security thresholds of individual attacks as well as collective attacks.} \label{Fig:Fig3}
\end{figure}

For each photon number $\mu_a$ and $\mu_b$, the protocol is then executed continuously and the basis reconciliation done in parallel to the data acquisition. In the reconciliation procedure the detections for compatible and incompatible basis choices by Alice and Bob are split. At this point all the incompatible detections are discarded. An important figure of merit for any QKD setup is the QBER, whose definition is $(N_i)/(N_c+N_i)$, where $N_c$ and $N_i$ are the correct and incorrect detections at Bob's detector. For this experiment the $N_c$ detections correspond to the sums of all the ones obtained when both MUB vectors chosen by Alice and Bob are the same. The $N_i$ detections are the sum of the detections when Alice and Bob do not choose the same MUB vectors. In the real-world scenario there will always be practical issues such as imperfect components and misalignment. Therefore, the detected probabilities will differ from the ideal ones. In order for any QKD protocol to be secure, the QBER must be below a certain value. This value depends on the attack strategy allowed for Eve, as well as the number of dimensions of the system. Following the bounds provided in \cite{Cerf_PRL_2002}, we calculate that with 16 dimensions the error rate threshold for secure key generation when considering individual attacks is $D^{\mathrm{ind}} = 37.50\%$. For the more general class of collective attacks we obtain an upper bound of $D^{\mathrm{coh}} = 28.97\%$. In our experimental implementation the measured QBER is shown in Fig.~3 together with the breakdown of the detections into the different types of detected events. In the case of using $\mu_a$, the average QBER is 13.4$\pm$4\%. In the QKD session with $\mu_b$, the average QBER is 15.6$\pm$7\%. One can note the stability provided by the setup due to the small variations in the QBER during the long experimental runs. In both cases the QBER is below the security thresholds of individual attacks as well as collective attacks. This indicates that a positive net key rate can be extracted from the sifted bits \cite{Cerf_PRL_2002}. Nevertheless, further theoretical studies are needed to optimize the optimum average photon number per pulse, as a function of the transmission distance for high-dimensional QKD. The average obtained sifted key rate is 27.0 and 11.1 bits / hour for $\mu_a$ and $\mu_b$, respectively. No post-processing was done in the sifted key bits. Finally, we have calculated the Shannon's entropy $H \equiv \sum\nolimits p_k\mathrm{log}p_k$ for the detected symbols for both $\mu_a$ and $\mu_b$, where $p_k$ is the probability of detecting state $k$. We obtain an entropy of $3.98\pm0.03$ and $3.92\pm0.05$ bits for the QKD sessions with $\mu_a$ and $\mu_b$, respectively. Both cases are close to the theoretical maximum of 4 bits/symbol when working in a 16-dimensional Hilbert space.

\section*{\label{sec:level1}Discussion}

Here, we have experimentally demonstrated an automated QKD session based on the BB84 protocol extended to 16-dimensional quantum states. The QBER remains constant and below the security limits, showing the robustness and stability of the setup. Our work therefore can be seen as an initial point for the development of real-world experimental high-dimensional QKD. It paves the way for future investigations in this area. For example, the new generation of SLMs based on ferroelectric liquid crystals allows for frame rates up to 1 kHz. So, one may envisage that they will allow for faster key generation rates in schemes similar to the setup presented here. Moreover, recent developments on integrated silicon photonics have demonstrated SLMs with a modulation speed of 150 MHz \cite{Qiu_scirep_2012}. They are designed to modulate the light in one transverse direction and, thus, they match the requirements for encoding information using the linear transverse momentum. Another line of investigation is the study of propagation of spatially-encoded qudits for QKD through optical fibers. This is challenging since single-mode fibers cannot be used and multi-mode fibers scramble the information across the modes. Nevertheless progress has been made recently using photonic crystal fibers \cite{Woerdman_PRL_11} and multi-mode fibers with active compensation of scattered light propagation \cite{Choi_PRL_2012}. This last technique is promising for our setup since it involves the transmission of images through multi-mode fibers. In principle, the image of Alice's SLM would be formed onto the fiber, propagated, then imaged again on Bob's SLM. Last, it is worth to mention that our scheme can also be adapted to investigate the performance of other QKD protocols in higher dimensions \cite{HengFan}, and study fundamental related concepts \cite{Gallego}.

\section*{Methods}

\subsection*{Mutually unbiased bases}
The two employed MUBs are displayed here. They obey the condition that $\mid \langle \phi^{(\alpha)}_{k} \mid \phi^{(\alpha')}_{k'} \rangle \mid^2 = \delta_{\alpha \alpha'} * \delta_{k k'} + \frac{1}{d} (1 - \delta_{\alpha \alpha'})$, where $\alpha$ and $\alpha'$ denote the two MUBs. $k$ and $k'$ label the states in a given MUB. For sake of simplicity, we describe the MUB's states in terms of unitary transformations. The MUBs are given by:

\begin{equation*}
U^{(\alpha)}=\frac{1}{\sqrt{16}}\left(\begin{array}{cccccccccccccccc} 1 & -1 & 1 & 1 & -1 & 1 & -1 & 1 & -1 & 1 & 1 & 1 & 1 & 1 & -1 & -1\\ -1 & 1 & 1 & 1 & -1 & 1 & 1 & -1 & 1 & 1 & -1 & -1 & 1 & 1 & 1 & -1\\ 1 & 1 & 1 & -1 & 1 & -1 & 1 & 1 & -1 & 1 & -1 & 1 & 1 & -1 & 1 & -1\\ 1 & 1 & -1 & 1 & 1 & 1 & 1 & -1 & 1 & -1 & 1 & 1 & 1 & -1 & -1 & -1\\ -1 & -1 & 1 & 1 & 1 & -1 & 1 & 1 & 1 & -1 & 1 & 1 & -1 & 1 & 1 & -1\\ 1 & 1 & -1 & 1 & -1 & 1 & 1 & 1 & -1 & -1 & -1 & 1 & -1 & 1 & 1 & 1\\ -1 & 1 & 1 & 1 & 1 & 1 & 1 & 1 & -1 & 1 & 1 & -1 & -1 & -1 & -1 & 1\\ 1 & -1 & 1 & -1 & 1 & 1 & 1 & 1 & 1 & -1 & -1 & -1 & 1 & 1 & -1 & 1\\ -1 & 1 & -1 & 1 & 1 & -1 & -1 & 1 & 1 & 1 & -1 & 1 & 1 & 1 & -1 & 1\\ 1 & 1 & 1 & -1 & -1 & -1 & 1 & -1 & 1 & 1 & 1 & 1 & -1 & 1 & -1 & 1\\ 1 & -1 & -1 & 1 & 1 & -1 & 1 & -1 & -1 & 1 & 1 & -1 & 1 & 1 & 1 & 1\\ 1 & -1 & 1 & 1 & 1 & 1 & -1 & -1 & 1 & 1 & -1 & 1 & -1 & -1 & 1 & 1\\ 1 & 1 & 1 & 1 & -1 & -1 & -1 & 1 & 1 & -1 & 1 & -1 & 1 & -1 & 1 & 1\\ 1 & 1
& -1 & -1 & 1 & 1 & -1 & 1 & 1 & 1 & 1 & -1 & -1 & 1 & 1 & -1\\ -1 & 1 & 1 & -1 & 1 & 1 & -1 & -1 & -1 & -1 & 1 & 1 & 1 & 1 & 1 & 1\\ -1 & -1 & -1 & -1 & -1 & 1 & 1 & 1 & 1 & 1 & 1 & 1 & 1 & -1 & 1 & 1 \end{array}\right),
\end{equation*}
\begin{equation*}
U^{(\alpha^\prime)}=\frac{1}{\sqrt{16}}\left(\begin{array}{cccccccccccccccc} 1 & 1 & 1 & 1 & 1 & 1 & 1 & 1 & 1 & 1 & 1 & 1 & 1 & 1 & 1 & 1\\ 1 & 1 & -1 & 1 & 1 & -1 & -1 & 1 & -1 & 1 & -1 & -1 & -1 & -1 & 1 & 1\\ 1 & -1 & 1 & 1 & -1 & -1 & 1 & -1 & 1 & -1 & -1 & -1 & -1 & 1 & 1 & 1\\ 1 & 1 & 1 & -1 & -1 & 1 & -1 & 1 & -1 & -1 & -1 & -1 & 1 & 1 & 1 & -1\\ 1 & 1 & -1 & -1 & 1 & -1 & 1 & -1 & -1 & -1 & -1 & 1 & 1 & 1 & -1 & 1\\ 1 & -1 & -1 & 1 & -1 & 1 & -1 & -1 & -1 & -1 & 1 & 1 & 1 & -1 & 1 & 1\\ 1 & -1 & 1 & -1 & 1 & -1 & -1 & -1 & -1 & 1 & 1 & 1 & -1 & 1 & 1 & -1\\ 1 & 1 & -1 & 1 & -1 & -1 & -1 & -1 & 1 & 1 & 1 & -1 & 1 & 1 & -1 & -1\\ 1 & -1 & 1 & -1 & -1 & -1 & -1 & 1 & 1 & 1 & -1 & 1 & 1 & -1 & -1 & 1\\ 1 & 1 & -1 & -1 & -1 & -1 & 1 & 1 & 1 & -1 & 1 & 1 & -1 & -1 & 1 & -1\\ 1 & -1 & -1 & -1 & -1 & 1 & 1 & 1 & -1 & 1 & 1 & -1 & -1 & 1 & -1 & 1\\ 1 & -1 & -1 & -1 & 1 & 1 & 1 & -1 & 1 & 1 & -1 & -1 & 1 & -1 & 1 & -1\\ 1 & -1 & -1 & 1 & 1 & 1 & -1 & 1 & 1 & -1 & -1 & 1 & -1 & 1 & -1 & -1\\ 1 & -1 & 1 & 1 & 1 & -1 & 1 & 1 & -1 & -1 & 1 & -1 & 1 & -1 & -1 & -1\\ 1 & 1 & 1 & 1 & -1 & 1 & 1 & -1 & -1 & 1 & -1 & 1 & -1 & -1 & -1 & -1\\ 1 & 1 & 1 & -1 & 1 & 1 & -1 & -1 & 1 & -1 & 1 & -1 & -1 & -1 & -1 & 1 \end{array}\right).
\end{equation*}

\subsection*{Acknowledgments}

The authors wish to thank A. Klimov for assistance with the generation of the MUBs, and J. Cari\~ne, A. Wolf and M. Figueroa for discussions related to the FPGAs. This work was funded by the grants FONDECYT 1120067, ICM P10-030F and CONICYT PFB08-024. G.B.X. acknowledges funding from FONDECYT 11110115. G.C. and E.S.G. acknowledge the financial support of CONICYT.


\begin{thebibliography}{999}

\bibitem{GPU1}
Murakami, T., Kasahara, R. \& Saito T.
An implementation and its evaluation of password cracking tool parallelized on GPGPU.
In \textit{International Symposium On Communication And Information Technologies (ISCIT)}, Tokyo Japan, Oct. 2010.

\bibitem{GPU2}
Vu, A.-D., Han, J.-I., Nguyen, H.-A., Kim, Y.-M. \& Im, Y.-M.
A homogeneous parallel brute force cracking algorithm on the GPU.
In \textit{International Conference On ICT Convergence (ICTC)}, Seoul, South Korea, Sept. 2011.

\bibitem{Gisin_RMP}
Gisin, N., Ribordy, G., Tittel, W. \& Zbinden, H.
Quantum cryptography.
\textit{Rev. Mod. Phys.} \textbf{74}, 145 (2002).

\bibitem{BB84}
Bennett, C. H. \& Brassard, G.
Quantum cryptography: public key distribution and coin tossing.
In \textit{Proceedings Of The IEEE International Conference On Computers, Systems And Signal Processing, Bangalore, India} (IEEE, New York, 1984), p. 175.

\bibitem{Wooters2}
Wootters, W. K. \& Fields, B. D.
Optimal state-determination by mutually unbiased measurements.
\textit{Ann. Phys.} \textbf{191}, 363--381 (1989).

\bibitem{Gobby_APL_2004}
Gobby, C., Yuan, Z. L. \& Shields, A. J.
Quantum key distribution over 122 km of standard telecom fiber.
\textit{Appl. Phys. Lett.} \textbf{84}, 3762 (2004).

\bibitem{Takesue_natphoton_2007}
Takesue, H. et al.
Quantum key distribution over a 40-dB channel loss using superconducting single-photon detectors.
\textit{Nature Photon.} \textbf{1}, 343 (2007).

\bibitem{Stucki_NJP_2009}
Stucki, D. et al.
High rate, long-distance quantum key distribution over 250 km of ultra low loss fibres.
\textit{New J. Phys.} \textbf{11}, 075003 (2009).

\bibitem{Guix_NJP_2009}
Xavier, G. B. et al.
Experimental polarization encoded quantum key distribution over optical fibres with real-time continuous birefringence compensation.
\textit{New J. Phys.} \textbf{11}, 045015, (2009).

\bibitem{Shields_APL_2010}
Dixon, A. R:, Yuan, A. L., Dynes, J. F., Sharpe, A. W. \& Shields, A. J.
Continuous operation of high bit rate quantum key distribution.
\textit{Appl. Phys. Lett.} \textbf{96}. 161102 (2010).

\bibitem{Guo_OL_2012}
Wang, S. et al.
2 GHz clock quantum key distribution over 260 km of standard telecom fiber.
\textit{Opt. Lett.} \textbf{37}, 1008 (2012).

\bibitem{Proakis_book}
Proakis, J. G. \& Salehi, M.
\textit{Digital Communications 5th Edition} (McGraw-Hill 2007).

\bibitem{Tittel_PRA_2000}
Bechmann-Pasquinucci, H. \& Tittel, W.
Quantum cryptography using larger alphabets.
\textit{Phys. Rev. A} \textbf{61}, 062308 (2000).

\bibitem{Boure_PRA_2001}
Bourennane, M., Karlsson, A. \& Bj\"{o}rk, G.
Quantum key distribution using multilevel encoding.
\textit{Phys. Rev. A} \textbf{64}, 012306 (2001).

\bibitem{Cerf_PRL_2002}
Cerf, N. J., Bourennane, M., Karlsson, A. \& Gisin, N.
Security of quantum key distribution using \textit{d}-level systems.
\textit{Phys. Rev. Lett.} \textbf{88}, 127902 (2002).

\bibitem{Zeilinger_NJP_2006}
Gr\"{o}blacher, S., Jennewein, T., Vaziri, A., Weihs, G. \& Zeilinger, A.
Experimental quantum cryptography with qutrits.
\textit{New J. Phys.} \textbf{8}, 75 (2006).

\bibitem{Howell_PRL_2007}
Ali-Kahn, I., Broadbent, C. J. \& Howell, J. C.
Large-alphabet quantum key distribution using energy-time entangled bipartite states.
\textit{Phys. Rev. Lett.} \textbf{98}, 060503 (2007).

\bibitem{Steve_PRL_2006}
Walborn, S. P., Lemelle, D. S., Almeida, M. P. \& Souto Ribeiro, P. H.
Quantum key distribution with higher-order alphabets using spatially encoded qudits.
\textit{Phys. Rev. Lett.} \textbf{96}, 090501 (2006).

\bibitem{Sheridan_PRA_2010}
Sheridan, L. \& Scarani, V.
Security proof for quantum key distribution using qudit systems.
\textit{Phys. Rev. A} \textbf{82}, 030301(R) (2010).

\bibitem{Twi1}
Grier, D. G.
A revolution in optical manipulation.
\textit{Nature} \textbf{424}, 21 (2003).

\bibitem{Padgett_2006}
Yao, E., Franke-Arnold, S. , Courtial, J., \& Padgett, M. J.
Observation of quantum entanglement using spatial light modulators.
\textit{Opt. Express} {\bf 14}, 13089-13094 (2006).

\bibitem{Glima_opex_2009}
Lima, G., Vargas, A., Neves, L., Guzm\'an, R. \& Saavedra, C.
Manipulating spatial qudit states with programmable optical devices.
\textit{Opt. Express}, \textbf{17}, 10688 (2009).

\bibitem{Padgett_nature_phys}
Dada, A. C., Leach, J., Buller, G. S., Padgett, M. J. \& Andersson, E.
Experimental high-dimensional two-photon entanglement and violations of generalized Bell inequalities.
\textit{Nature Phys.} \textbf{7}, 677 (2011).

\bibitem{Torres_nature_phys}
Hendrych, M. et al.
Experimental estimation of the dimension of classical and quantum systems. \textit{Nature Phys.} \textbf{8}, 588 (2012).

\bibitem{Glima_opex_2011}
Lima, G. et al.
Experimental quantum tomography of photonic qudits via mutually unbiased basis.
\textit{Opt. Express} \textbf{19}, 3542 (2011).

\bibitem{SPadua_opex}
Pimenta, W. et al.
Minimal state tomography of spatial qubits using a spatial light modulator.
\textit{Opt. Express} {\bf 18}, 24423 (2010).

\bibitem{Willner_nat_photon}
Wang, J. et al.
Terabit free-space data transmission employing orbital angular momentum multiplexing.
\textit{Nature Photon.} \textbf{6}, 488 (2012).

\bibitem{Neves_PRL_2005}
Neves, L. et al.
Generation of Entangled States of Qudits using Twin Photons.
\textit{Phys. Rev. Lett.} \textbf{94}, 100501 (2005).

\bibitem{Moreno_opex}
Moreno, I., Vel\'aquez, P., Fern\'andez-Pousa, C. R. \& S\'anchez-L\'opez, M. M.
Jones matrix method for predicting and optimizing the optical modulation properties of a liquid-crystal display.
\textit{J. Appl. Phys.} \textbf{94}, 3697Ð3702 (2003).

\bibitem{Klimov}
Romero, J. L.,  Bj\"{o}rk, G., Klimov, A. B., \& S\'{a}nchez-Soto, L. L.
Structure of the sets of mutually unbiased bases for N qubits.
\textit{Phys. Rev. A} \textbf{72}, 062310 (2005).

\bibitem{Lo_phase_random}
Zhao, Y., Bing, Q. \& Lo, H.-K.
Experimental quantum key distribution with active phase randomization.
\textit{Appl. Phys. Lett.} \textbf{90}, 044106 (2007).

\bibitem{Qiu_scirep_2012}
Qiu, C., Chen, J., Xia, Y. \& Xu, Q.
Active dielectric antenna on chip for spatial light modulation.
\textit{Sci. Rep.} \textbf{2}, 855 (2012).

\bibitem{Woerdman_PRL_11}
L\"{o}ffler, W. et al.
Fiber Transport of Spatially Entangled Photons.
\textit{Phys. Rev. Lett.} \textbf{106}, 240505 (2011).

\bibitem{Choi_PRL_2012}
Choi, Y. et al.
Scanner-free and wide-Field Endoscopic Imaging by Using a Single Multimode Optical Fiber.
\textit{Phys. Rev. Lett.} \textbf{109}, 203901 (2012).

\bibitem{HengFan}
Xiong, Z.-X. et al.
General quantum key distribution in higher dimension.
\textit{Phys. Rev. A} \textbf{85}, 012334 (2012).

\bibitem{Gallego}
Gallego, R., Brunner, N., Hadley, C., \& Ac\'{i}n, A.
Device-Independent Tests of Classical and Quantum Dimensions.
\textit{Phys. Rev. Lett.} \textbf{105}, 230501 (2010).

\end{thebibliography}
\end{document}